\documentclass[twocolumn,amsmath,amssymb,prb,showpacs,longbibliography]{revtex4-1}
\usepackage[english]{babel}
\usepackage{graphicx}
\usepackage{dcolumn}
\usepackage{bm}
\usepackage[usenames,dvipsnames]{xcolor}
\usepackage{stackengine}
\usepackage[colorlinks=true,linkcolor=blue,citecolor=blue,urlcolor=blue]{hyperref}
\usepackage[
  group-digits = false
]{siunitx}
\usepackage{makecell}
\usepackage{multirow}
\usepackage{dsfont}
\usepackage{soul}
\usepackage{braket}
\usepackage{upgreek}
\usepackage{pgfplots}
\usepackage{cellspace}
\usepackage{overpic,picture}
\pgfplotsset{compat=newest}
\usetikzlibrary{pgfplots.groupplots}
\usetikzlibrary{arrows}
\usetikzlibrary{decorations.text}

\newcommand{\MR}[1]{\mathrm{#1}}
\newcommand{\ud}{\mathrm{d}}
\newcommand{\iu}{\mathrm{i}}
\newcommand{\Tr}{\mathrm{Tr}}
\DeclareMathOperator{\IM}{Im}
\DeclareMathOperator{\RE}{Re}

\newcommand{\VEC}[1]{\mathbf{#1}}

\newcommand{\MUB}{\mu_{\text{B}}}

\begin{document}
\setstcolor{red}
\title{Comparative study of methodologies to compute the intrinsic Gilbert damping: interrelations, validity and physical consequences}

\author{Filipe S. M. Guimar\~aes}
\email{f.guimaraes@fz-juelich.de}
\affiliation{Peter Gr\"unberg Institut and Institute for Advanced Simulation, Forschungszentrum J\"ulich \& JARA, 52425 J\"ulich, Germany}

\author{J. R. Suckert}
\affiliation{Peter Gr\"unberg Institut and Institute for Advanced Simulation, Forschungszentrum J\"ulich \& JARA, 52425 J\"ulich, Germany}

\author{Jonathan Chico}
\affiliation{Peter Gr\"unberg Institut and Institute for Advanced Simulation, Forschungszentrum J\"ulich \& JARA, 52425 J\"ulich, Germany}

\author{Juba Bouaziz}
\affiliation{Peter Gr\"unberg Institut and Institute for Advanced Simulation, Forschungszentrum J\"ulich \& JARA, 52425 J\"ulich, Germany}

\author{Manuel dos Santos Dias}
\affiliation{Peter Gr\"unberg Institut and Institute for Advanced Simulation, Forschungszentrum J\"ulich \& JARA, 52425 J\"ulich, Germany}

\author{Samir Lounis}
\affiliation{Peter Gr\"unberg Institut and Institute for Advanced Simulation, Forschungszentrum J\"ulich \& JARA, 52425 J\"ulich, Germany}

\date{\today}

\begin{abstract}

Relaxation effects are of primary importance in the description of magnetic excitations, leading to a myriad of methods addressing the phenomenological damping parameters.
In this work, we consider several well-established forms of calculating the intrinsic Gilbert damping within a unified theoretical framework, mapping out their connections and the approximations required to derive each formula.
This scheme enables a direct comparison of the different methods on the same footing and a consistent evaluation of their range of validity.
Most methods lead to very similar results for the bulk ferromagnets Fe, Co and Ni, due to the low spin-orbit interaction strength and the absence of the spin pumping mechanism.
The effects of inhomogeneities, temperature and other sources of finite electronic lifetime are often accounted for by an empirical broadening of the electronic energy levels.
We show that the contribution to the damping introduced by this broadening is additive, and so can be extracted by comparing the results of the calculations performed with and without spin-orbit interaction.
Starting from simulated ferromagnetic resonance spectra based on the underlying electronic structure, we unambiguously demonstrate that the damping parameter obtained within the constant broadening approximation diverges for three-dimensional bulk magnets in the clean limit, while it remains finite for monolayers.
Our work puts into perspective the several methods available to describe and compute the Gilbert damping, building a solid foundation for future investigations of magnetic relaxation effects in any kind of material.

\end{abstract}

\maketitle

\section{Introduction}

Dynamical processes lie at the core of magnetic manipulation.
From the torques acting on the magnetic moments to how fast they relax back to their equilibrium orientations, a material-specific time-dependent theory is essential to describe and predict their behavior.
In most cases, the description of the time evolution of the magnetization is done via micromagnetics\cite{Wei:2012fp} or atomistic spin dynamics (ASD)\cite{Etz:2015kn,Eriksson:2017dc} approaches, in which the magnetization is considered either as a classical continuous vector field or as individual 3D vectors on a discrete lattice, respectively.
They have been successfully used to describe a plethora of magnetic phenomena, ranging from spin waves in low dimensional magnets\cite{Bergqvist:2013iu}, domain walls\cite{Schieback:2007fd} and skyrmion\cite{Iwasaki:2013hb} dynamics to thermal stability of magnetic textures\cite{vonMalottki:2017cr}. These approaches model the magnetization dynamics via a phenomenological equation of motion that contains both precessional and relaxation terms.

A first attempt to address these processes was performed by Landau and Lifshitz (LL), by considering a Larmor-like precessional torque and adding to it a (weaker) damping term of relativistic origin~\cite{Landau:1935tr}. 
Since its phenomenological inception in 1935, the precise nature of the relaxation processes has been a source of intense debate. 
In particular, the original LL formulation was found to not properly describe situations in which the damping was large. 
This problem was addressed by Gilbert, who introduced a Rayleigh-like dissipation term into the magnetic Lagrangian, thus obtaining the now-ubiquitous Landau-Lifshitz-Gilbert (LLG) equation~\cite{Gilbert:2004gx},
\begin{equation}\label{llg}
\begin{split}
\frac{\ud\VEC{M}}{\ud t} &= -\gamma\,\VEC{M}\times\VEC{B} + \frac{\alpha}{M}\,\VEC{M}\times \frac{\ud\VEC{M}}{\ud t} \\
&= -\widetilde{\gamma}\,\VEC{M}\times\VEC{B} - \frac{\alpha\widetilde{\gamma}}{M}\,\VEC{M}\times \left(\VEC{M}\times\VEC{B}\right) \ .
\end{split}
\end{equation}
where $\gamma>0$ is the gyromagnetic factor, $\VEC{M}$ is the (spin) magnetic moment, $\VEC{B}$ is the time-dependent effective magnetic field acting on $\VEC{M}$, and $\alpha$ is the scalar damping parameter named after Gilbert.
The upper form of the LLG equation is due to Gilbert, and the lower one shows that it is equivalent to a LL equation with a renormalized gyromagnetic factor, $\widetilde{\gamma} = \gamma/(1+\alpha^2)$.
The first term in the right-hand side of Eq.~\eqref{llg} describes the precession of the magnetic moments around the effective field, while the second term is the Gilbert damping one, that describes the relaxation of the magnetic moments towards $\VEC{B}$.
This equation corrects the previously mentioned issue for large values of $\alpha$, for which the original LL equation is expected to fail~\cite{Kikuchi:1956ed,Mallinson:1987jq}. 

The ferromagnetic resonance (FMR) technique is one of the most common procedures to probe magnetization dynamics~\cite{Heinrich:1966da}, in which the damping parameter is related to the linewidth of the obtained spectra~\cite{Kittel:1948ur}.
Although many measurements have been carried out in bulk materials\cite{Rodbell:1964ey,Heinrich:1966da,Bhagat:1966ih,Bhagat:1974iu,Heinrich:1979bda,Cochran:1998cu}, their description at low temperatures is still controversial \cite{Hankiewicz:2007iy,Mankovsky:2013ii,Costa:2015hp,Edwards:2016be}. 
This can be attributed to the different intrinsic and extrinsic mechanisms that can contribute to the relaxation processes~\cite{Sanders:1974ke,Schoen:2017hj,Cadwell:1996hp,Hickey:2009tj,Mondal:2016bb,Kittel:1958cv,Streib:2018ef,Holanda:2018hr,Volkenshtein:1973bl,Tserkovnyak:2002ju,Korenman:1972gx,Vollmer:2003io,Arias:1999uk,Lindner:2003uu}.
When varying the temperature, two distinct regimes could be identified in the measured relaxation parameters~\cite{Kambersky:1976gi}.
For high temperatures, a proportionality between the linewidth and the temperature was observed in most of the experiments.
It was called resistivity-like, due to the similarity with the temperature dependence of this quantity. 
A conductivity-like regime (linewidth inversely proportional to the temperature) was identified at low temperatures for certain materials such as Ni~\cite{Bhagat:1966ih,Heinrich:1979bda}, but not for Fe~\cite{Bhagat:1968in,Cochran:1998cu}. 
It was also seen that different concentrations of impurities affected this low-temperature regime, even suppressing it altogether~\cite{Bhagat:1974iu}.

From the theoretical point-of-view, the calculation of the Gilbert parameter is a challenging problem due to the many different mechanisms that might be at play for a given material~\cite{Prange:1972kp,Bland:2005gx}.
Perhaps this is why most of the theoretical approaches have focused on contributions to the damping from electronic origin.
The ultimate goal then becomes the development of a predictive theory of the Gilbert damping parameter, based on the knowledge of a realistic electronic structure of the target magnetic material.
The ongoing efforts to complete this quest have resulted in the development of a myriad of techniques~\cite{Kambersky:1970hf,Kambersky:1976gi,brattserkovbauer,Garate:2009ic,Costa:2015hp,Edwards:2016be}.
Comparisons between a few of these approaches are available~\cite{Sakuma:2015ex,Turek:2015bw}, including experimental validation of some methods~\cite{Schoen:2016gc,Schoen:2017hj}, but a complete picture is still lacking.

\begin{figure*}[htb]
    \centering
    \hypersetup{pdfborder=0 0 0}
    \begin{overpic}[
      width=1.0\textwidth]{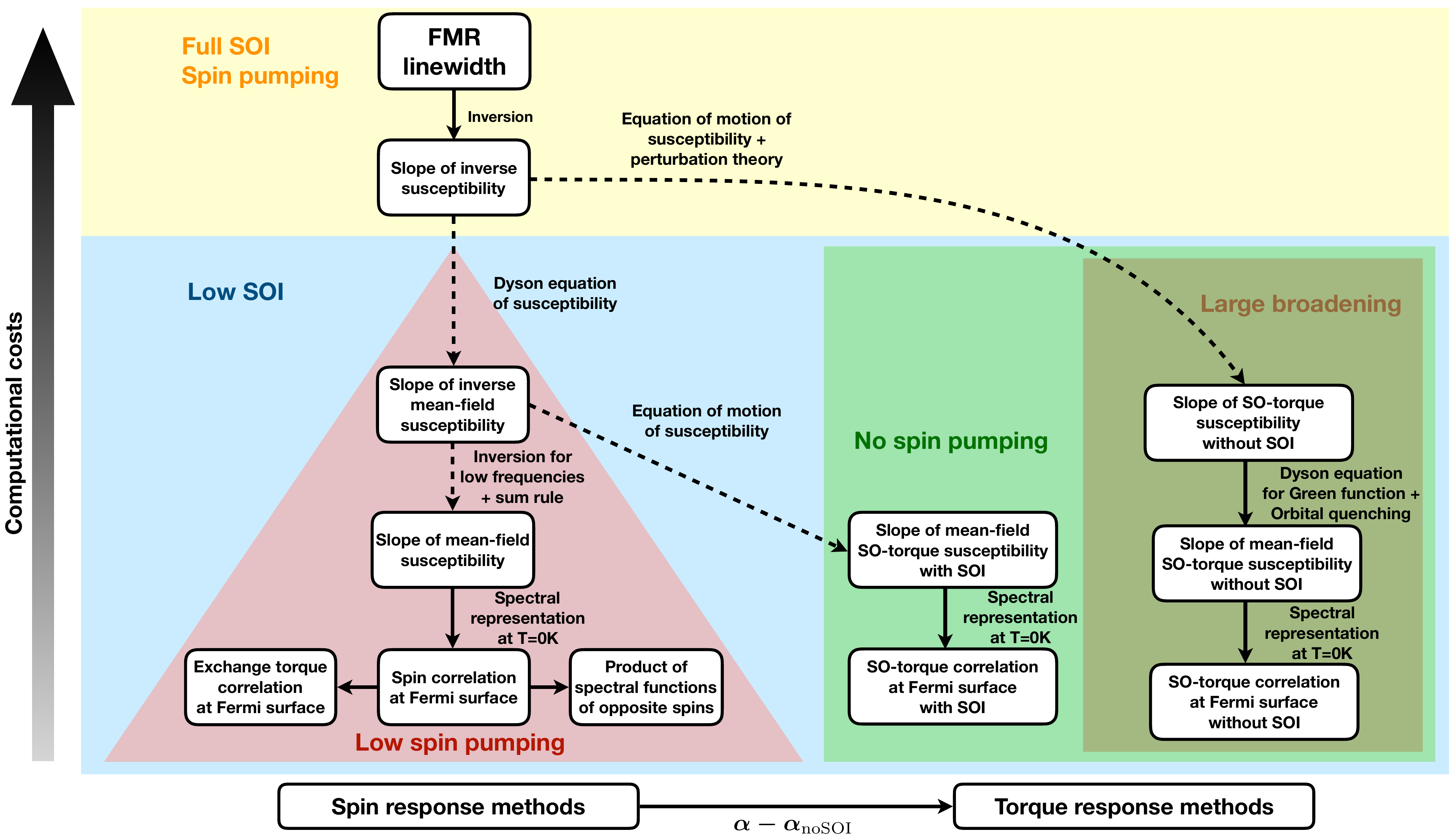}
      \put(26.3,51.6){\hyperref[sec:FMR]{\makebox(51pt,24pt)[lb]{}}}%
      \put(26.3,43){\hyperref[chiinv]{\makebox(51pt,24pt)[lb]{}}}%
      \put(26.3,27.5){\hyperref[chi0inv]{\makebox(51pt,24pt)[lb]{}}}%
      \put(25.8,17.6){\hyperref[slopeofchi0withoutSOI]{\makebox(55pt,24pt)[lb]{}}}%
      \put(26.3,8.2){\hyperref[xctcm]{\makebox(51pt,24pt)[lb]{}}}%
      \put(13.1,8.2){\hyperref[xctcm]{\makebox(51pt,24pt)[lb]{}}}%
      \put(39.4,8.2){\hyperref[xctcm]{\makebox(51pt,24pt)[lb]{}}}%
      \put(58.5,17.6){\hyperref[tsm]{\makebox(71pt,24pt)[lb]{}}}%
      \put(58.5,8.2){\hyperref[sotcm]{\makebox(71pt,24pt)[lb]{}}}%
      \put(78.8,26.3){\hyperref[tsm]{\makebox(70pt,24pt)[lb]{}}}%
      \put(79.2,16.6){\hyperref[tsm]{\makebox(71pt,24pt)[lb]{}}}%
      \put(79,7.2){\hyperref[sotcm]{\makebox(71pt,24pt)[lb]{}}}%
      \put(19.5,1){\hyperref[spinmethods]{\makebox(123pt,14pt)[lb]{}}}%
      \put(65.5,1){\hyperref[ttresponse]{\makebox(124pt,14pt)[lb]{}}}%
    \end{overpic}
    \caption{Diagram exhibiting the different methods investigated in this work, their connections and range of validity. 
    Two groups are identified: one related to the spin susceptibility (spin response methods), including the ferromagnetic resonance and the slope of the inverse susceptibility that involves a direct mapping of this quantity to the LLG equation; and the other associated with torque responses, for which approximations need to be taken.
    The steps indicated by solid lines represent exact connections, while dashed arrows involve some kind of approximation. 
    The arrow on the left points from the methods that require less computational power (lower part) to the more demanding ones (upper part).
    Boxes are hyper-linked with the respective equations and sections.
\label{fig:diagram}}
\end{figure*}
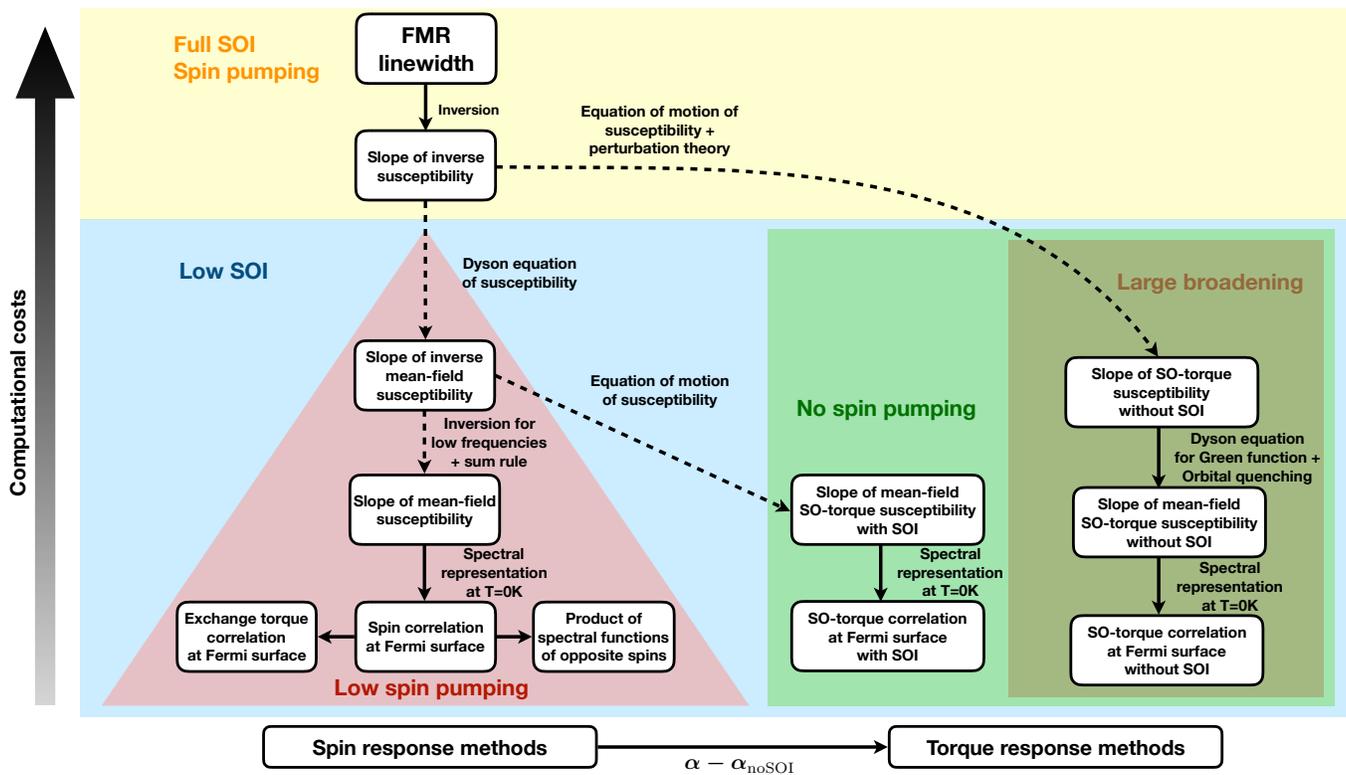

We clarify this subject by addressing most of the well-established methods to calculate the Gilbert damping from first principles.
First, we connect the many different formulas, highlighting the approximations made in each step of their derivations, determining what contributions to the damping they contain, and establishing their range of validity.
These are schematically illustrated in Fig.~\ref{fig:diagram}.
Second, we select a few approaches and evaluate the Gilbert damping within a unified and consistent framework, making use of a multi-orbital tight-binding theory based on first-principles electronic structure calculations.
FMR simulations and the mapping of the slope of the inverse susceptibility are used to benchmark the torque correlation methods based on the exchange and spin-orbit torques.
We apply these different techniques to bulk and monolayers of transition metals (Fe, Ni and Co), for which the spin pumping mechanism is not present and only the spin-orbit interaction (SOI) contributes to the relaxation.
Disorder and temperature effects are included by an empirical broadening of the electronic energy levels~\cite{Kambersky:1976gi,Garate:2009ic,Thonig:2014kt,Freimuth:2014kq}.
Third, we engage a longstanding question regarding the behavior of the damping in the low-temperature and low-disorder limits: should the intrinsic contribution to the Gilbert damping diverge for clean systems?
Our results using the constant broadening model demonstrate that the divergence is present in the clean limit of 3D systems but not of the 2D ones~\cite{Barati:2014gh}, which we attest by eliminating the possibility of them being caused by numerical convergence issues or different anisotropy fields. 
Our results also indicate that the limit $\omega\rightarrow 0$ is not responsible for the divergence of the intrinsic damping, as it is commonly attributed~\cite{Kambersky:1976gi,Hankiewicz:2007iy,Garate:2009ic,Freimuth:2017eu}.
Finally, we propose a new way to obtain the spin-orbit contribution that excludes the fictitious temperature/disorder contribution caused by the artificial broadening~\cite{Wieser:2015jt,Mahfouzi:2017kv}: they can be discounted by subtracting the values of damping calculated without SOI.
For bulk systems, this yields the total damping, while in layered materials this method should also discount part of the spin pumping contribution.
In Ref.~\onlinecite{Mankovsky:2013ii}, where temperature and disorder are included via a CPA analogy, a similar artificial increase of $\alpha$ for high temperatures was removed by including vertex corrections.

This work is organized as follows. We start, in Sec.~\ref{overview}, with a brief overview of the different methods proposed in the literature.
In Sec.~\ref{theory}, we explain the theory used to calculate the response functions.
We then turn to the distinct theoretical forms of calculating the damping: In Sec.~\ref{spinmethods}, we analyze the different approaches related to the spin-spin responses, while in Sec.~\ref{ttresponse}, the torque methods are explored.
We then discuss the obtained results and conclude in Sec.~\ref{sec:conclusions}.
The Hamiltonian used in the microscopic theory is given in Appendix~\ref{appx:theory}, while the anisotropy fields for the 3D and 2D systems together with the transverse dynamical magnetic susceptibility given by the LLG equation are given in Appendix~\ref{appx:llg}.

\section{Overview of methods addressing intrinsic Gilbert damping}
\label{overview}

We now focus on the different methods to describe the microscopic contributions to the Gilbert parameter, which encompasses effects that transfers energy and angular momentum out of the magnetic system.
Within these mechanisms, the relativistic SOI comes to the fore.
This is often referred to as the intrinsic contribution to the damping, and was first identified by Landau and Lifshitz~\cite{Landau:1935tr}.
The origin of this damping mechanism lies in the non-hermiticity of the relativistic corrections to the spin Hamiltonian when the magnetization precesses~\cite{Hickey:2009tj,Mondal:2016bb}.
The elementary magnetic excitations, called magnons, can also be damped via Stoner excitations (electron-hole pairs with opposite spins)~\cite{Korenman:1972gx,Mook:1973cf,Vollmer:2003io}.
Alternatively, the conduction electrons can carry spin angular momentum even in absence of the SOI.
This leads to damping via the spin-pumping mechanism~\cite{Tserkovnyak:2002ju,Simanek:2003wx,Mills:2003ub,Liu:2014fa}.

Early models proposed to describe these processes already argued that the interaction between the magnetic moments and the conduction electrons is a key ingredient~\cite{Heinrich:1967fma}.
This led to the so-called breathing Fermi surface model, where the shape of the Fermi surface depends on the orientation of the magnetization through the SOI~\cite{Kambersky:1970hf}.
This approach, however, could only capture the conductivity-like regime, which diverges at low temperatures.
The decay of magnons into Stoner excitations was also considered early on~\cite{Prange:1972kp}, describing well the experimental behavior of Ni but also missing the increase at larger temperatures of other materials.

An important progress was made by Kambersky using the spin-orbit torque correlation function to calculate the damping parameter~\cite{Kambersky:1976gi}.
This approach captures both conductivity- and resistivity-like behaviors, which were shown to originate from the intra- and interband transitions, respectively~\cite{Gilmore:2007ev}.
Recently, this so-called torque correlation method was re-obtained using a different perturbative approach~\cite{Hankiewicz:2007iy}, spurring discussions about the validity of the obtained results, specially the divergence caused by the intraband transitions~\cite{Edwards:2016be}.
A similar method also based on torque correlation functions was developed using a scattering theory approach~\cite{brattserkovbauer} involving the exchange torque operator instead of the spin-orbit torque one.
Results obtained in this way also present diverging behavior in the clean limit of 3D structures~\cite{Mankovsky:2013ii}.
A similar scattering framework was used to explain the enhancement of the Gilbert damping due to the spin pumping in thin films\cite{Tserkovnyak:2002ju}.
Yet another method relating the Gilbert damping to the spin-spin response was proposed and related to the existing spin-orbit torque correlation method~\cite{Garate:2009ic}.
It also presented diverging intraband contributions when the parameter used to broaden the delta functions (which mimics the effect of disorder or temperature) was taken to zero~\cite{Garate:2009jx}.
The vertex corrections proposed in Ref.~\onlinecite{Garate:2009jx} did not remove this divergence.
More recently, Costa and Muniz~\cite{Costa:2015hp} showed that the damping parameters of layered structures remain finite in the zero broadening limit, when extracted directly from the linewidth of the dynamical magnetic susceptibility (within the random phase approximation).

Several of these methods have been implemented for material-specific calculations~\cite{Kunes:2002ia,Gilmore:2007ev,Starikov:2010cz,Ebert:2011gx,Mankovsky:2013ii,Barati:2014gh,Thonig:2014kt}, and some approaches were compared and related~\cite{Garate:2009ic,Sakuma:2015ex,Turek:2015bw,Durrenfeld:2015dc}.
In this work, we start our analysis with the uniform frequency-dependent spin-spin susceptibility, which is measured experimentally in FMR setups, to derive the other expressions for the damping parameter based on the spin- and torque-correlation methods.

\section{Microscopic Theory}
\label{theory}

We begin by setting the grounds of the theory we use to evaluate the different formulas of the Gilbert damping on equal footing.
The electronic structure of the system is described by the mean-field Hamiltonian
\begin{equation}\label{hamilt}
\begin{split}
\hat{H} = \hat{H}_0+\hat{H}_\text{xc}+\hat{H}_\text{SOI}+\hat{H}_\text{ext}
\end{split} \  .
\end{equation}
The paramagnetic band structure is described by $\hat{H}_0$ within a multi-orbital tight-binding parametrization.
An effective local electron-electron interaction within the mean-field approximation is included in $\hat{H}_\text{xc}$, which is responsible for ferromagnetism.
We also account for spin-orbit interaction through $\hat{H}_\text{SOI}$, and the interaction with external static magnetic fields via $\hat{H}_\text{ext}$.
The explicit forms of all the terms are given in Appendix~\ref{appx:theory}.

In this work, we investigate the different methods to compute the intrinsic Gilbert damping utilizing the prototypical bulk magnets Fe (bcc), Co (fcc) and Ni (fcc), and also square lattices corresponding to the (001) planes of those materials, with the same nearest-neighbor distances as in its bulk forms.

For simplicity, we consider the spin-orbit interaction and the local effective Coulomb interaction only on the $d$ orbitals, with $U=\SI{1}{\electronvolt}$~\cite{Janak:1977ce,Himpsel:1991ea,Sasoglu:2011ch} for all systems, and the spin-orbit strengths $\lambda_\text{SOI}^\text{Fe}=\SI{54}{\milli\electronvolt}$~\cite{Guimaraes:2015fl}, $\lambda_\text{SOI}^\text{Co}=\SI{70}{\milli\electronvolt}$~\cite{Thonig:2017jf}, and $\lambda_\text{SOI}^\text{Ni}=\SI{133}{\milli\electronvolt}$~\cite{Thonig:2017jf}.
The magnetic ground state is found by self-consistently enforcing charge neutrality for the bulk materials~\cite{Costa2010}.
For the monolayer cases, the total number of electrons in the atomic plane is decreased to $n=7.3$ (Fe), $n=8.1$ (Co) and $n=9.0$ (Ni), as the remaining charge spills into the vacuum (which we are not explicitly taking into account within the model).
The ground-state properties (spin moment $M$, orbital moment $M_\ell$ and magnetic anisotropy energy $K$) obtained within this framework are listed in Table~\ref{table:gs}.
The easy axis for all the bulk systems and the monolayers were found to be along the (001) direction.
We emphasize that our goal is not to achieve the most realistic description of the electronic structure of these materials, but rather to define a concrete set of cases that allow us to compare the different methods to compute the Gilbert damping.

\begin{table}[b]
  \renewcommand*{\arraystretch}{1.2}
  \begin{center}
  \begin{ruledtabular}
  \begin{tabular}{ c | c | c | c | c | c | c }
       & \multicolumn{3}{c|}{bulk} & \multicolumn{3}{c}{monolayer}\\
       \hline
       & bcc Fe                          &  fcc Co    & fcc Ni    & Fe & Co & Ni\\
       \hline
        $M (\MUB)$                       & 2.32 & 1.48 & 0.43 & 2.90 & 1.90 & 0.96  \\
        $M_{\ell} (\MUB)$                & 0.072 & 0.079 & 0.055 & 0.28 & 0.22 & 0.20  \\
        $K (\SI{}{\milli\electronvolt})$  & 0.19 & 0.26 & 0.084 & 1.7 & 1.8 & 1.9  \\
  \end{tabular}
  \end{ruledtabular}
  \caption{Ground state properties of the investigated systems. $M$ and $M_\ell$ denotes the spin and orbital magnetic moments, respectively. Values obtained for $\eta=\SI{1.36}{\milli\electronvolt}$. The magnetic anisotropy constant $K$ is obtained from the anisotropy fields given by Eq.~\eqref{maefields}.
  \label{table:gs}}
  \end{center}
\end{table}

The magnetic excitations are described using linear response theory, where the transverse magnetic response $\delta\VEC{M}(t)$ due to an oscillatory magnetic field $\delta\VEC{B}(t)$ is given by~\cite{Kubo:1957cl}
\begin{equation}
\begin{split}
\delta M^\alpha(t) =  \!\!\int\!\MR{d}t'\,\chi^{\alpha\beta}(t-t')\,\delta B^\beta(t')
\end{split} \  ,
\end{equation}
where the convention to sum over repeated indices of the components $\beta=\{x,y,z\}$ is used.
This approach captures the orbitally-averaged part of the response.
The magnetic susceptibility is given by
\begin{equation}\label{chi}
\begin{split}
\chi^{\alpha\beta}\left(t-t'\right) &= -4\,\big\langle\big\langle \hat{S}^\alpha(t),\hat{S}^\beta(t')\big\rangle\big\rangle \\
&= 4\iu\,\big\langle \big[\hat{S}^\alpha\left(t\right),\hat{S}^\beta\left(t'\right) \big]\big\rangle
\end{split} \  ,
\end{equation}
in atomic units. 
$\hat{S}^\alpha\left(t\right)$ is the $\alpha$-component of the spin operator.
In the first line of the equation above, we reprise the double-bracket notation of Zubarev for the spin-spin retarded Green function~\cite{Zubarev:1960vo}.
This notation is convenient for the derivations of Sec. \ref{ttresponse}.

For the crystal symmetries of the systems we are interested in, it is convenient to work in the circular basis $\hat{S}^\pm = \hat{S}^x\pm \iu\hat{S}^y$, which diagonalizes the susceptibility matrix with components $\chi^{-+}(t)$ and $\chi^{+-}(t)$.
The frequency- and wave vector-dependent transverse susceptibility $\chi^{-+}(\mathbf{q},\omega)$ is obtained within the random phase approximation (RPA), which captures the collective spin wave modes~\cite{Mills:1967hh,Costa:2015hp}, as well as the possible decay into particle-hole excitations (Stoner modes) described by the single-particle response function $\chi_0^{-+}(\omega)$.
Considering matrices that take into account the orbital dependency, the two susceptibilities are related by
\begin{equation}\label{dyson}
\begin{split}
[\underline{\chi}^{-+}]^{-1} = [\underline{\chi}^{-+}_0]^{-1} - \frac{1}{4}\underline{U}\ .
\end{split}
\end{equation}
Here, $\underline{U}_{\mu\nu}=U\delta_{\mu\nu}$ is a matrix with the effective local Coulomb interaction strength within the $d$ orbitals.
It plays a similar role to the exchange-correlation kernel in the adiabatic local-density approximation of time-dependent DFT calculations~\cite{Lounis2010}.
We define the transverse magnetic response of the system by summing the susceptibility matrix over all the $d$ orbitals.

The uniform single particle transverse susceptibility $\underline{\chi}_0^{-+}(\omega) = \underline{\chi}_0^{-+}(\VEC{q}=0,\omega)$, obtained within the mean-field approximation, is expressed in terms of the single-particle Green functions as
\begin{equation}\label{chi0}
\begin{split}
\underline{\chi}_{0,\mu\nu}^{-+}(\omega) = \frac{1}{\pi N}\sum_\VEC{k}&\int^{\epsilon_{\text{F}}}\!\!\!\!\!\ud\varepsilon\;\left\{G^{\uparrow\uparrow}_{\mu\nu}(\VEC{k},\varepsilon+\omega)\,\IM\,G^{\downarrow\downarrow}_{\nu\mu}(\VEC{k},\varepsilon)\right.\\
+&\left.\,\IM\,G^{\uparrow\uparrow}_{\mu\nu}(\VEC{k},\varepsilon)\,\left[G^{\downarrow\downarrow}_{\mu\nu}(\VEC{k},\varepsilon-\omega)\right]^* \right\}\  .
\end{split}
\end{equation}
The sum is over the wave vectors in the first Brillouin zone, with $N$ their number.
The indices $\mu,\nu$ represent orbitals and $\epsilon_{\text{F}}$ is the Fermi level.

In the spirit of many preceding works~\cite{Kambersky:1976gi,Garate:2009ic,Thonig:2014kt,Freimuth:2014kq},
the effect of temperature and disorder is modeled by introducing a constant band broadening $\eta$ on the energy levels, such that $G(\omega) \rightarrow G(\omega + \iu\eta)$.
The imaginary part of the Green function is then defined as $\IM\,G_{\mu\nu}(\omega) = \frac{1}{2\iu} \left\{G_{\mu\nu}(\omega+\iu\eta) -G_{\mu\nu}(\omega-\iu\eta)\right\}$.
This approach attempts to capture all the intrinsic effects originated from the electronic structure of the system.

The imaginary part of the susceptibility is related to the energy dissipation of the system~\cite{Callen:1951be}, encoding the relaxation mechanism of the magnetization towards equilibrium.
The damping parameter is then obtained by mapping the transverse magnetic susceptibility obtained from the quantum mechanical calculation described above to the phenomenological form provided by the LLG, Eq.~\eqref{llg}.
On the following sections, we present different mapping procedures involving several approximations and explore their range of validity when the broadening $\eta$ is taken to zero (clean limit).

\section{Spin response methods}
\label{spinmethods}

\subsection{Ferromagnetic resonance}
\label{sec:FMR}

\begin{figure*}[ht!]
    \centering
        \includegraphics[width=1.0\textwidth]{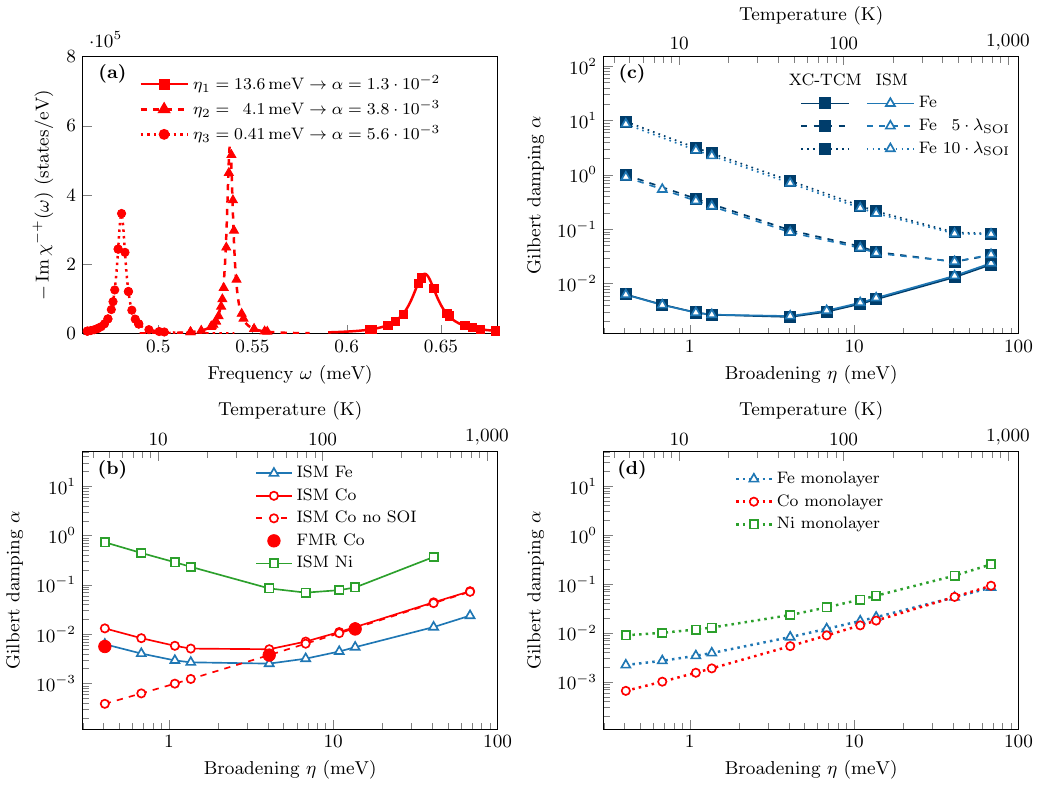}
    \caption{Characteristics of the Gilbert damping in 3D and 2D systems in presence and absence of SOI. 
	(a) Ferromagnetic resonance spectra for fcc Co, in presence of spin-orbit interaction and no external field, calculated for three different decreasing broadenings $\eta_1=\SI{13.6}{\milli\electronvolt}$ (solid), $\eta_2=\SI{4.1}{\milli\electronvolt}$ (dashed) and $\eta_3=\SI{0.41}{\milli\electronvolt}$ (dotted). The values of the Gilbert damping given in the legend box, obtained by fitting to Eq.~\eqref{chifmr}, decrease from the first case to the second, but increases again when $\eta$ is further decreased.
	(b) Gilbert damping in presence of spin-orbit interaction for bcc Fe (blue triangles), fcc Co (red circles, solid line) and fcc Ni (green squares) as a function of the broadening, obtained from the slope of the inverse susceptibility, Eq.~\eqref{chiinv}. All values were computed with $10^8$ k-points in the full Brillouin zone.
	Solid red circles are the values obtained from the FMR spectra in (a), while the open red circles connected by dashed lines represent the damping parameter for fcc Co when SOI is not included in the calculations.
	(c) Damping parameter for bcc Fe for different SOI strenghts: $\lambda_\text{SOI} = \SI{54.4}{\milli\electronvolt}$, $5\times\lambda_\text{SOI}$, and $10\times\lambda_\text{SOI}$.
	(d) Gilbert damping of Fe, Co and Ni monolayers in the presence of SOI. No increase in the Gilbert damping is seen when the broadening $\eta$ is decreased.
\label{fig:alpha}}
\end{figure*}

Magnetic excitations can be investigated by applying time-dependent perturbations.
This is done in FMR experiments where the magnetic sample is subjected to a static magnetic field and an oscillatory radio-frequency one.
By varying either the strength of the static component or the frequency of the oscillatory field, the system can be driven through magnetic resonance.
This setup yields the uniform mode of the transverse magnetic susceptibility.
As the Gilbert parameter describes the relaxation mechanisms of the magnetization, it is related to the linewidth of the resonance peak~\cite{Costa:2015hp,IbanezAzpiroz:2017gc}.

We simulate this kind of experiments by calculating the transverse magnetic response relying on the linear response theory discussed in Sec.~\ref{theory}, and mapping the imaginary part of the susceptibility into the result obtained from the LLG equation (see Appendix~\ref{appx:llg}),
\begin{equation}\label{chifmr}
\IM\chi^{-+}(\omega) = -\frac{2\alpha \gamma \omega M}{[\omega-\gamma \left(B_{\text{ext}}+B_{\text{an}}\right)]^2+(\alpha\omega)^2}\  .
\end{equation}
When fixing the frequency and varying $B_{\text{ext},z}$, this function presents a resonance at $B_\text{res} = (\omega-\gamma B_{\text{an},z})/\gamma$ with linewidth given by the full width at half maximum $\Delta B = 2\alpha\omega/\gamma$.
On the other hand, when the field is kept fixed and the frequency is varied, the resonance is located at $\omega_\text{res} = \gamma (B_{\text{ext},z}+B_{\text{an},z})/\sqrt{1+\alpha^2}$ with full width at half maximum approximately given by $\Delta \omega \approx 2\alpha\gamma|B_{\text{ext},z}+B_{\text{an},z}|$, in the limit $\alpha\ll 1$\cite{IbanezAzpiroz:2017gc}.

The Gilbert parameter can then be obtained either by fitting Eq.~\eqref{chifmr} or through the ratio between the linewidth and the resonance position.
In this sense, a divergence of the damping when $\eta\rightarrow0$ seems counter-intuitive, since this would imply that either the resonance position ($B_\text{res}$ or $\omega_\text{res}$) goes to zero or that the corresponding linewidth increases drastically.
In the presence of SOI, the SU(2) rotational symmetry is broken and the anisotropy field $B_{\text{an},z}$ shifts the resonance position to a finite value --- it costs a finite amount of energy to set the magnetization into precession~\cite{dosSantosDias:2015bh}.
Therefore, the divergence of the damping parameter can only happen if the linewidth increases and goes to infinity.

To verify this claim, we simulate FMR experiments in fcc Co bulk by calculating the imaginary part of the transverse magnetic susceptibility as a function of the frequency $\omega$, in the presence of the spin-orbit interaction.
In Fig.~\ref{fig:alpha}a, we present the obtained spectra for different values of the broadening $\eta$.
When a relatively large value of the broadening is used, $\eta=\SI{13.6}{\milli\electronvolt}$ (solid curve), the spectra displays a broad resonance peak, which can be characterized by a value $\alpha=\SI{1.3e-2}{}$, obtained by fitting the linear response data with Eq.~\eqref{chifmr}.
When the broadening of the energy levels is decreased to $\eta=\SI{4.1}{\milli\electronvolt}$ (dashed curve), the peak shifts and becomes sharper ($\alpha=\SI{3.8e-3}{}$), as one intuitively expects when disorder and/or temperature decreases. 
Notice that most of the change in $\alpha$ is due to the change in the peak width, while the resonance shift is relatively small.
This can be viewed as a consequence of the smaller energy overlap between the bands, which decrease possible interband transitions~\cite{Gilmore:2007ev}.
Surprisingly, by further decreasing the broadening to $\eta=\SI{0.41}{\milli\electronvolt}$ (dotted curve), the peak becomes broader when compared to the previous case, with $\alpha=\SI{5.6e-3}{}$.
This counter-intuitive result represents a shorter lifetime of the magnetic excitation when the electronic lifetime (mean time between two successive scattering events) $\tau = \eta^{-1}$ becomes longer.

Obtaining the damping from the FMR curves is computationally demanding, though.
The response function must be calculated for many frequencies (or magnetic fields) to resolve the peak.
For the case of low broadenings that require many k-points in the Brillouin zone for a converged result, this task becomes prohibitive.
In the next section, we provide alternative methods to obtain the Gilbert parameter based on the static limit of the susceptibility, and compare their outcomes with the ones obtained using the resonance approach.

\subsection{Inverse Susceptibility Method}
\label{staticmethods}

We proceed now to investigate a different mapping of the microscopic transverse susceptibility to the LLG equation and possible approximations to simplify the calculation of the Gilbert damping.
From Eq.~\eqref{llg+-}, one can see that $\alpha$ defines the slope of the imaginary part of the inverse susceptibility~\cite{Garate:2009ic}, i.e.,
\begin{equation}\label{chiinv}
\alpha = 2\gamma M \lim_{\omega\rightarrow0} \frac{\IM[\chi^{-+}(\omega)]^{-1}}{\omega} \  .
\end{equation}
We will refer to this as the {\em inverse susceptibility method} (ISM).
The mapping to the LLG model of the slope at small frequencies has a great advantage over the FMR one since it only requires a single frequency-point calculation, instead of a full sweep over frequencies or magnetic fields for the fitting procedure.

In Fig.~\ref{fig:alpha}b, we display the damping parameter for bcc Fe, fcc Co and fcc Ni bulk systems calculated as a function of the electronic energy broadening.
We also include the results obtained from the FMR approach (solid symbols), which compare well with the ISM given in Eq.~\eqref{chiinv}.
Note that although Eq.~\eqref{chiinv} has an explicit linear dependence on the spin moment $M$, the susceptibility implicitly depends on its value.
The obtained curves are inversely related to $M$: highest for Ni ($M \sim 0.45\MUB$), lowest for Fe ($M \sim 2.3\MUB$) and Co in-between ($M \sim 1.5\MUB$).
This trend is confirmed by setting the SOI strength $\lambda_\text{SOI}$ to the same values for all the elements (not shown).
The position of the minimum value of $\alpha$ is connected with $\lambda_\text{SOI}$, which determines when the intraband or interband transitions become more important\cite{Gilmore:2007ev}.
To substantiate this claim, we employed the technique of artificially scaling the $\lambda_\text{SOI}$, as previously done in connection to the magnetic anisotropy energy\cite{Halilov:1998hs}.
The results are shown in Fig.~\ref{fig:alpha}c, where the SOI strength $\lambda_\text{SOI}$ of Fe bulk is magnified by factors of 5 and 10.
Indeed, the minimum can clearly be seen to shift to larger values of $\eta$.

An important aspect to be considered is the convergence of Eq.~\eqref{chi0} --- failing to achieve numerical precision may give rise to spurious results~\cite{Steiauf:2005bv,Barati:2014gh}.
This can be partly solved using sophisticated schemes to perform those calculations~\cite{Muniz:2002jw,Lounis:2011bn}.
When the broadening is lowered, the convergence of the wave vector summation is affected by the increasingly dominant role of the poles of the Green functions in the vicinity of the Fermi energy.
For that reason, to capture the intricacies of the electronic states --- in particular, the important contributions from the small gaps opened by the weak SOI ---, we calculated the slope of the response function using a very fine integration mesh on the Brillouin zone reaching up to $10^9$ k-points.
The results in Fig.~\ref{fig:alpha}c also demonstrate that the divergence is not an issue of numerical convergence, since this behavior is shifted to larger values of broadenings, for which the convergence is more easily achieved.

Nevertheless, such diverging effect only occurs in the presence of spin-orbit interaction.
In Fig.~\ref{fig:alpha}b we also display the values of $\alpha$ for Co fcc obtained using the ISM when the SOI is not included in the calculations (circles connected by dashed lines).
In this case, $\alpha_\text{noSOI}$ linearly goes to zero when the broadening is decreased~\cite{Costa:2015hp}.
The non-vanishing damping when SOI is not present can be interpreted as originating from the finite electronic lifetimes introduced by the constant broadening parameter. 
As it stands, $\eta$ represents the coupling to a fictitious reservoir~\cite{Wieser:2015jt,Mahfouzi:2017kv} providing dissipation mechanisms that physically should originate from disorder or temperature, for example.

Obtaining the damping from the FMR spectra when SOI is not present requires an applied magnetic field, such that the resonance frequency becomes finite and avoiding an infinite response at zero frequency (representing no cost of energy due to the rotational symmetry, i.e., the Goldstone mode).
Nevertheless, the results presented in Fig.~\ref{fig:alpha}d were obtained using the ISM without any applied field.
Calculations with an applied magnetic field shifting the peak to the original anisotropy energy were indistinguishable from those values (with variations smaller than $3\%$).
This is accordance to the phenomenological expectations expressed through Eq.~\eqref{llg+-}, where the slope is independent of the magnetic field.

One can put our results for bulk ferromagnets into perspective by comparing with low dimensional systems.
We investigated this case within our linear response approach, using monolayers of Fe, Co and Ni.
The calculations follow the same procedure, except that the sum over $\VEC{k}$ vectors in Eq.~\eqref{chi0} is restricted to the 2D Brillouin zone.
The results are presented as triangles (Fe), circles (Co) and squares (Ni) connected by dotted lines in Fig.~\ref{fig:alpha}d, and once again exhibit a monotonous decay with the decrease of $\eta$.
We note that previous calculations of the damping parameter in thin films also did not find it to increase rapidly for decreasing broadening~\cite{Barati:2014gh,Costa:2015hp}.

Besides the dimensionality, another main difference from the bulk to the layered case is the larger anisotropy fields of the latter (see Table~\ref{table:gs}).
Nevertheless, this cannot explain the non-diverging behavior in the monolayers.
We have already shown that by artificially increasing the SOI strength of the bulk --- and, consequently, its anisotropy field ---, the conductivity-like behavior of the damping occurs at even larger broadenings (see Fig.~\ref{fig:alpha}c).
On the other hand, to rule out a possible divergence happening at lower broadenings ($\eta<\SI{0.1}{\milli\electronvolt}$, not reachable in our calculations), we have also scaled up $\lambda_\text{SOI}$ of the monolayers by one order of magnitude.
This resulted in larger dampings, nonetheless, the same decreasing behaviour with $\eta\rightarrow 0$ was observed (not shown).
Therefore, the divergence can only be attributed to the three-dimensionality of the ferromagnet.

\subsection{Approximate static limit methods}

We now look back to Fig.~\ref{fig:diagram} and proceed to perform approximations on Eq.~\eqref{chiinv} in order to simplify the calculations of the damping parameter.
Here we follow Ref.~\onlinecite{Garate:2009ic}.
First, we use Eq.~\eqref{dyson} that relates the RPA susceptibility matrix to the mean-field response matrix $\underline{\chi}_0$, such that $\IM\,\chi^{-1}\approx \IM\,\chi_0^{-1}$.
Although $\underline{U}$ is a real matrix, the sum over orbitals ($\chi = \sum_{\mu\nu}\underline{\chi}_{\mu\nu}$) ends up mixing the real and imaginary parts of the matrix elements.
Only when $\RE\,\chi_0^{-1} = U/4$ the relation above becomes an equality.
This means that, within our model with $U$ acting only on the $d$ orbitals, $\chi$ must also be defined by summing over those orbitals only.
Under the previous assumption, we obtain
\begin{equation}
\label{chi0inv}
\alpha \approx 2\gamma M \lim_{\omega\rightarrow0} \frac{\IM[\chi^{-+}_0(\omega)]^{-1}}{\omega} \  .
\end{equation}
This relation is only valid when $\chi^{-+}_0$ is decoupled from the other types of susceptibilities (transverse and longitudinal), as in the systems we investigate in this work.
The damping parameter can therefore be obtained from the single-particle transverse susceptibility $\chi_0$.

For frequencies $\omega$ in the meV range (where the collective spin excitations are located), $\chi_0^{-+}$ has a simple $\omega$-dependence~\cite{Lounis:2015ho}:
\begin{equation}
\label{chi0exp}
\chi^{-+}_0(\omega) \approx \RE\,\chi_0(0) + \iu\,\omega\,\IM\,\chi'_0(0)\  .
\end{equation}
where $\chi'_0(0)=\left.\frac{\ud\chi^{-+}_0}{\ud \omega}\right|_{\omega=0}$.
These results are valid also in the presence of spin-orbit coupling.
Using Eq.~\eqref{chi0exp}, the Gilbert damping can be written as
\begin{equation}
\label{chi0expinv2}
\begin{split}
\alpha
\approx& -2\gamma M\left[\RE\,\chi^{-+}_0(0)\right]^{-2}\lim_{\omega\rightarrow0} \frac{\IM\,\chi^{-+}_0(\omega)}{\omega}
\end{split} \  .
\end{equation}
Although the expansion of the susceptibility for low frequencies was used, no extra approximation is employed, since Eq.~\eqref{chi0inv} is calculated in the limit $\omega\rightarrow0$.
$\RE\,\chi^{-+}_0(0)$ can be obtained using the sum rule that relates the static susceptibility with the magnetic moments~\cite{dosSantosDias:2015bh}.
For $3d$ transition metals, the external and the spin-orbit fields are three orders of magnitude smaller than $U$, and so the static susceptibility of the bulk systems reads $\RE\chi^{-+}_0(0) \approx 4/U$.
Thus,
\begin{equation}\label{slopeofchi0withoutSOI}
\begin{split}
\alpha \approx& -\frac{\gamma MU^2}{8} \lim_{\omega\rightarrow0} \frac{\IM\chi^{-+}_0(\omega)}{\omega}
\end{split} \  .
\end{equation}

Finally, from Eq.~\eqref{chi0} it is possible to show that Eq.~\eqref{slopeofchi0withoutSOI} simplifies as
\begin{equation}\label{xctcm}
\begin{split}
\alpha
&\approx \frac{\gamma MU^2}{2\pi N} \sum_\VEC{k,\mu\nu}\Tr\{\IM\,G_{\nu\mu}(\VEC{k},\epsilon_F)\,\hat{S}^-\,\IM\,G_{\mu\nu}(\VEC{k},\epsilon_F)\,\hat{S}^+\}\\
&= \frac{\gamma}{2M \pi N} \sum_\VEC{k,\mu\nu} \Tr\{\IM\,G_{\nu\mu}(\VEC{k},\epsilon_F)\,\hat{T}^-_\text{xc}\,\IM\,G_{\mu\nu}(\VEC{k},\epsilon_F)\,\hat{T}^+_\text{xc}\}\\
&= \frac{\gamma MU^2\pi}{8 N} \sum_\VEC{k,\mu\nu} n^\downarrow_{\nu\mu}(\VEC{k},\epsilon_F)\,n^\uparrow_{\mu\nu}(\VEC{k},\epsilon_F)
\end{split} \  .
\end{equation}
where $n^\sigma_{\mu\nu}(\VEC{k},\epsilon_F) = -\frac{1}{\pi}\IM\,G^{\sigma\sigma}_{\mu\nu}(\VEC{k},\epsilon_F)$ is the matrix element of the spectral function of spin $\sigma$ calculated at $\VEC{k}$ and $\epsilon_F$.
The second equation is written in terms of the ``exchange-correlation torque operator'', $T^{\pm}_\text{xc} = -\iu\,\big[\hat{S}^\pm,\hat{H}_\text{xc}\big]=\mp\iu\,UM \hat{S}^\pm$.
This result is equivalent to the one obtained in Ref.~\onlinecite{brattserkovbauer}, which we reference as the {\em exchange torque correlation method} (XC-TCM) --- although, in reality, it relates $\alpha$ with the spin-spin response.
The last step in Eq.~\eqref{xctcm} connects the damping with the product of spectral functions of opposite spins at the Fermi level, as shown theoretically in Ref.~\onlinecite{Lounis:2015ho} and confirmed experimentally in Ref.~\onlinecite{Schoen:2016gc}.

In Fig.~\ref{fig:alpha}c, we compare the results obtained with this approximated method with the ISM described before, for the different values of SOI scalings.
For the bulk transition metals we investigate, the approximation is very good, since the SOI is relatively small.
In fact, even when the SOI is scaled one order of magnitude higher, the results of the XC-TCM are still very good.

The formulas in Eq.~\eqref{xctcm} show that we have arrived at the bottom of the triangle in Fig.~\ref{fig:diagram}.
These forms do not involve an integral over energy, which simplifies substantially the calculation of $\alpha$.
For that reason, they are suitable for first-principles approaches (e.g., Refs.~\onlinecite{Ebert:2011gx,Mankovsky:2013ii}).
This concludes our investigations of the spin response methods.
In the next section, we take a different path to calculate the Gilbert damping.

\section{Torque response methods}
\label{ttresponse}

Despite the simplicity of the methods based on the spin susceptibility discussed in the previous section, seminal work was based on a different type of response function.
The main idea, first proposed by Kambersky~\cite{Kambersky:1976gi}, is to directly relate $\alpha$ to the spin-orbit interaction.
Here, our aim is twofold.
First, we connect the spin susceptibility with the spin-orbit torque response via the equation of motion, clarifying the damping mechanisms captured by this formalism.
Second, we compare the results obtained with both types of methods.

We start with the equation of motion for the spin-spin susceptibility.
Its time-Fourier transform can be written as~\cite{Hankiewicz:2007iy}
\begin{equation}\label{eqmov1}
\omega\,\big\langle\big\langle \hat{S}^-,\hat{S}^+\big\rangle\big\rangle_\omega
= M + \big\langle\big\langle \big[\hat{S}^-,\hat{H}\big],\hat{S}^+\big\rangle\big\rangle_\omega
\  ,
\end{equation}
where $M= -2\,\big\langle\hat{S}^z\big\rangle$.
From the Hamiltonian given in Eq.~\eqref{hamilt}, the commutator $[\hat{S}^-,\hat{H}\big]$ has four contributions: kinetic (spin currents, from $\hat{H}_0$), exchange torque (from $\hat{H}_\text{xc}$), external torque (from $\hat{H}_\text{ext}$) and spin-orbit torque (from $\hat{H}_\text{SOI}$).
In presence of SOI, the total spin magnetic moment is not a conserved quantity and spin angular momentum can be transferred to the orbital degrees of freedom.
For bulk systems subjected to static external fields and in the present approximation for the electron-electron interaction, the only two non-vanishing torques are due to the external field and the spin-orbit interaction.
It also follows from these assumptions that the mechanisms that contribute to the relaxation arises then from the spin-orbit torques $\hat{T}^\pm_\text{SOI} = -\iu\,\big[\hat{S}^\pm,\hat{H}_\text{SOI}\big]$ and from the broadening of the energy levels $\eta$.

It can be shown~\cite{Hankiewicz:2007iy} that the inverse of the susceptibility $\chi^{-+}(\omega) = \big\langle\big\langle \hat{S}^-,\hat{S}^+\big\rangle\big\rangle_\omega$ is given by
\begin{equation}\label{eqmovinv}
\begin{split}
\big[\chi^{-+}(\omega)\big]^{-1} &= \big[\chi^{-+}_\text{noSOI}(\omega)\big]^{-1} \big[1 + \chi^{-+}_\text{noSOI}(\omega)\,\Gamma(\omega)\big]^{-1}\\
&\approx \big[\chi^{-+}_\text{noSOI}(\omega)\big]^{-1} - \Gamma(\omega)
\end{split} \  .
\end{equation}
Here, $\chi^{-+}_\text{noSOI}(\omega)$ is the susceptibility calculated excluding the SOI contribution to the Hamiltonian.
The connection between the two susceptibilities in Eq.~\eqref{eqmovinv} is provided by the quantity
\begin{equation}
\begin{split}
M^2\,\Gamma(\omega) = \iu\,\big\langle\big[\hat{T}^-_\text{SOI},\hat{S}^+\big]\big\rangle + \big\langle\big\langle \hat{T}^-_\text{SOI},\hat{T}^+_\text{SOI}\big\rangle\big\rangle_\omega
\end{split} \  .
\end{equation}
Using Eq.~\eqref{chiinv}, and noticing that the first term on the right-hand side of the equation above does not contribute to the imaginary part, we find
\begin{equation}\label{tsm}
\alpha = \alpha_\text{noSOI} - \frac{2\gamma}{M} \lim_{\omega\rightarrow0} \frac{\IM\,\big\langle\big\langle \hat{T}^-_\text{SOI},\hat{T}^+_\text{SOI}\big\rangle\big\rangle_\omega}{\omega} \  .
\end{equation}
$\alpha_\text{noSOI}$ is the contribution obtained by inputting $\chi^{-+}_\text{noSOI}(\omega)$ into Eq.~\eqref{chiinv}, which is finite due to the broadening $\eta$.

Kambersky~\cite{Kambersky:1976gi} first obtained this same result following a different approach.
In our framework, this would involve starting from Eq.~\eqref{dyson} and exploiting the consequences of the fact that the collective spin excitations ($\omega\sim\SI{}{\milli\electronvolt}$) have low frequencies when compared to the exchange energy ($U \sim\SI{}{\electronvolt}$).
On the other hand, Hankiewicz \textit{et al.}~\cite{Hankiewicz:2007iy} described the same expansion for low SOI, and justified its use for $\omega\lesssim\gamma B_\text{ext}$.
Finally, Edwards~\cite{Edwards:2016be} shows that this formula is equivalent to a perturbation theory correct to $\lambda_\text{SOI}^2$ (compared to $\gamma B_\text{ext}-\omega$).
For that reason, he suggests that the states used in the calculation of $\big\langle\big\langle \hat{T}^-_\text{SOI},\hat{T}^+_\text{SOI}\big\rangle\big\rangle_\omega$ should not include SOI, since the operator $\hat{T}^-_\text{SOI} \propto \lambda_\text{SOI}$.
Due to the orbital quenching in the states without SOI, this leads to the absence of intraband contributions and, consequently, of the divergent behavior for $\eta\rightarrow0$\cite{Umerski:2017cn}.

In this approach, temperature and disorder effects are included in $\alpha_\text{noSOI}$ (shown in Fig.~\ref{fig:alpha}d for Co), while the spin-orbit intrinsic broadening is calculated by the second term in Eq.~\eqref{tsm}, which can also be obtained as $\alpha - \alpha_\text{noSOI}$.
An extra advantage of calculating the damping as the aforementioned difference is that one explicitly subtracts the contributions introduced by $\eta$, providing similar results to those obtained with vertex corrections~\cite{Mankovsky:2013ii}.
Considering the torque-torque response within the mean-field approximation (an exact result in the perturbative approach\cite{Edwards:2016be}), we obtain, similarly to Eq.~\eqref{xctcm},
\begin{equation}\label{sotcm}
\begin{split}
\alpha - &\alpha_\text{noSOI}
= \\
&\frac{2\gamma}{M \pi N} \sum_\VEC{k} \Tr\{\IM\,\underline{G}(\VEC{k},\epsilon_F)\,\hat{T}^-_\text{SOI}\,\IM\,\underline{G}(\VEC{k},\epsilon_F)\,\hat{T}^+_\text{SOI}\}
\end{split} \  .
\end{equation}
In this formula, the involved quantities are matrices in spin and orbital indices and the trace runs over both.
This is known as Kambersy's formula, commonly used in the literature~\cite{Gilmore:2007ev,Garate:2009ic,Barati:2014gh,Thonig:2014kt,Sakuma:2015ex}, which we refer to as {\em spin-orbit torque correlation method} (SO-TCM).
As in Eq.~\eqref{xctcm}, it relates the damping to Fermi level quantities only.
When the SOI is not included in the calculation of the Green functions $G(\VEC{k},\epsilon_F)$ and enters only through the torque operators, we name it {\em perturbative} SO-TCM~\cite{Edwards:2016be}.
These methods are placed at the bottom right of Fig.~\ref{fig:diagram}, with the main approximations required indicated by the long dashed arrows.

\begin{figure}
    \centering
	\includegraphics[width=1.0\columnwidth]{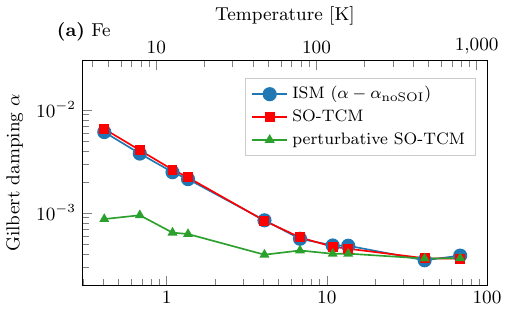}
       \\ \includegraphics[width=1.0\columnwidth]{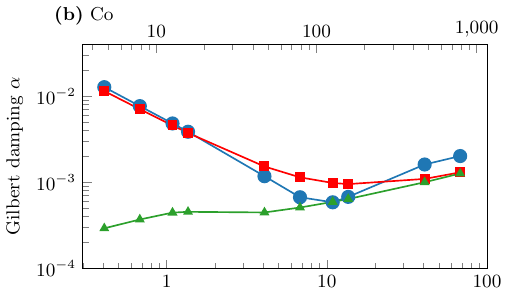}
       \\ \includegraphics[width=1.0\columnwidth]{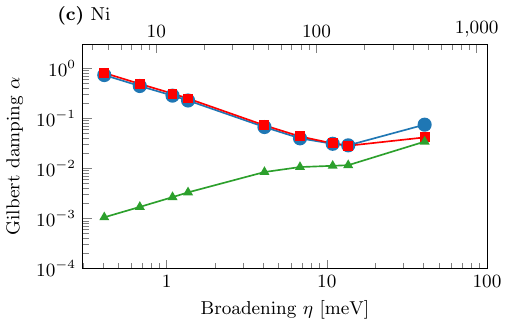}
    \caption{Comparison between $\alpha - \alpha_\text{noSOI}$ for (a) Fe bcc, (b) Co fcc and (c) Ni fcc, obtained using the inverse susceptibility method (ISM) with the spin-orbit-torque correlation method (SO-TCM) with and without SOI in the states (perturbative SO-TCM). All the points were computed with $10^8$ k-points in the full Brillouin zone.
}
    \label{fig:comparison}
\end{figure}

We now proceed to compare these approaches with the ISM explained in Sec.~\ref{staticmethods}.
Fig.~\ref{fig:comparison} presents the calculations of the SOI contribution to the damping parameter of bulk Fe (a), Co (b) and Ni (c) using the SO-TCM obtained in Eq.~\eqref{sotcm}, when no external field is applied.
Both approaches, including SOI (red curve with squares) in the Green functions or not (green curve with triangles), are shown.
For a meaningful comparison, we compute $\alpha - \alpha_\text{noSOI}$ within the ISM.

We first note that the perturbative approach suggested by Edwards~\cite{Edwards:2016be} describes reasonably well the large broadening range (i.e., mostly given by the interband transitions), but deviates from the other approaches for low $\eta$.
This is an expected behaviour since it does not include the intraband transitions that display the $\eta^{-1}$ behavior within the constant broadening model.
In the clean limit, the Gilbert damping computed from the perturbative SO-TCM approaches zero for all elements, in a very monotonic way for Co and Ni, but not for Fe.
This method is thus found to be in agreement with the other ones only when $\lambda_\text{SOI}\ll\eta$.
The SO-TCM formula including the SOI in the states (i.e., Kambersky's formula) matches very well $\alpha$ obtained within ISM in the whole range of broadenings. 

\begin{figure}[ht!]
    \centering
        \includegraphics[width=1.0\columnwidth]{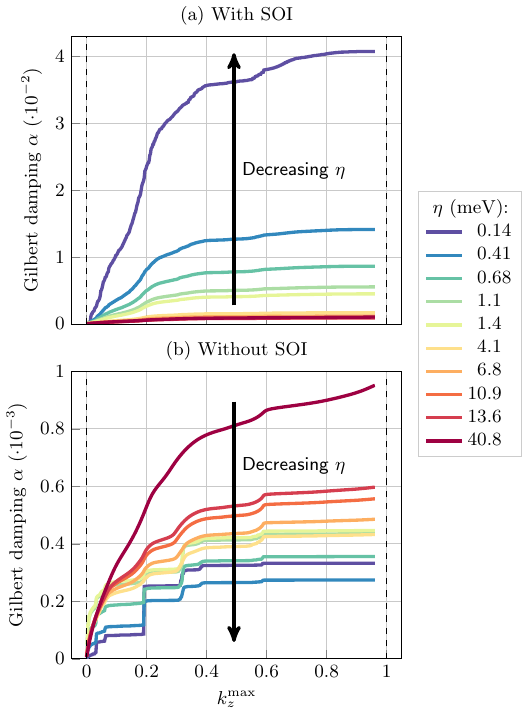}
    \caption{Integrated Gilbert damping for fcc Co as a function of $k_z$ plotted against the maximum value, $k_z^\text{max}$ (see Eq.~\eqref{eq:ialpha}), with SOI (a) and without SOI (b). The curves were obtained using the SO-TCM given in Eq.~\eqref{sotcm}. Colors represent different values of the broadening $\eta$ (in units of $\SI{}{\milli\electronvolt}$). The value of $\alpha$ for $k_z^\text{max}=0$ (i.e., a single value of $k_z$ in the sum) represents a two-dimensional system, whilst for $k_z^\text{max}=1$ the sum covers the whole 3D Brillouin zone. In the latter case, the damping decreases when $\eta$ is decreased without SOI, while it increases drastically when SOI is present. For 2D systems, the 
     \label{fig:ialpha}
}
\end{figure}
Finally, after demonstrating that the SO-TCM provides very similar results to the ISM, we can use it to resolve the wave-vector-dependent contributions to the Gilbert parameter by planes in the reciprocal space, as
\begin{equation}\label{eq:ialpha}
\begin{split}
\alpha(k_z^\text{max}) &= \sum_{|k_z|}^{k_z^\text{max}} \alpha(k_z)
\end{split} \  ,
\end{equation}
where $\alpha(k_z)$ is given by the right-hand side of Eq.~\eqref{sotcm} summed over $k_x,k_y$.
The result, displayed in Fig.~\ref{fig:ialpha}, uses 100 million k-points for all curves and shows the expected divergence in presence of SOI and a decrease with $\eta$ when this interaction is absent.
In every case, most of the contribution arises from the first half ($k_z^\text{max}<0.4$).
Note that when the broadening of the energy levels is low, the integrated alpha without SOI (Fig.~\ref{fig:ialpha}b) displays step-like contributions, while when SOI is present, they are smoother.
This is a consequence of the damping being caused by interband transitions in the former and intraband in the latter.

\begin{figure}
    \centering
        \includegraphics[width=1.0\columnwidth]{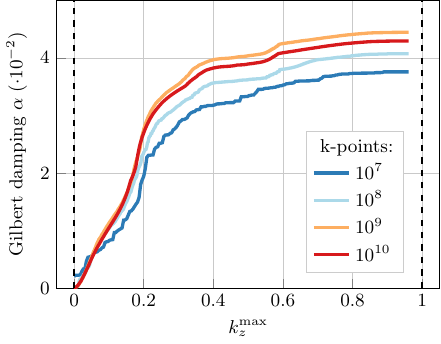}
    \caption{Integrated Gilbert damping for fcc Co as a function of $k_z$ plotted against the maximum value, $k_z^\text{max}$, for $\eta = \SI{0.14}{\milli\electronvolt}$ and different amount of k-points (up to 10 billion) in the Brillouin zone.
     \label{fig:ialpha_kpt}
}
\end{figure}
The convergence of the previous results for the smallest $\eta$ including SOI were tested with respect to the total number of k-points in the Brillouin zone in Fig.~\ref{fig:ialpha_kpt}.
By going from 10 million to 10 billion k-points, the results vary $\sim20\%$.
However, compared with the result shown in Fig.~\ref{fig:ialpha}a, the damping gets even larger, corroborating once more the divergent results.

\section{Discussions}
\label{sec:discussions}

In this section, we make a few final remarks on the previously obtained results and we go beyond bulk systems to comment on the approximations taken and additional physical mechanisms that may come into play in other materials.
We also provide a new analytical explanation for the divergence of the damping parameter within the constant broadening model.

Our first comment regards the application of static magnetic fields $B$.
As described in Refs.~\onlinecite{Hankiewicz:2007iy,Edwards:2016be}, the approximations done in Eq.~\eqref{eqmovinv} to derive an expression for $\alpha$ involves comparisons between the excitation energy and $B$.
However, all the results we have presented here were obtained in absence of static fields.
We also performed calculations including external magnetic fields up to $B\sim\SI{7}{\tesla}$, and the computed damping parameter is weakly influenced by their presence. 
We conclude that the validity of the SO-TCM formula given in Eq.~\eqref{sotcm} does not hinge on having a magnetic field, supporting the arguments already given in Ref.~\onlinecite{Hankiewicz:2007iy}.

A further remark concerns the approximations made to obtain the mean-field result in Eq.~\eqref{slopeofchi0withoutSOI}.
We assumed that SOI is weak when using the magnetic sum rule.
This approximation may break down when this is not the case.
The spin pumping also affects the magnetic sum rule, which may worsen the agreement with the ISM results.
Although this contribution is not present in the investigated (bulk-like) systems, it plays an important role in magnetic multilayers.
This effect enhances the damping factor~\cite{Tserkovnyak:2002ju,Simanek:2003wx,Mills:2003ub}.
Furthermore, the SO-TCM explicitly excludes spin pumping, as this is described by $\hat{I}^-_\text{S} = -\iu\,\big[\hat{S}^-,\hat{H}_0\big]$, dropped from the equation of motion.
These validity conditions are indicated in Fig.~\ref{fig:diagram} by the large blue rectangle (low SOI), red triangle (low spin pumping) and green rectangle (no spin pumping).

Another mechanism that opens new spin relaxation channels is the coupling between transverse and longitudinal excitations induced by the SOI.
This was one of the reasons raised in Ref.~\onlinecite{Costa:2015hp} to explain the divergence of the damping parameter.
However, this is absent not only when the system has full spin rotational symmetry~\cite{Muniz:2003vs}, but also when rotational symmetry is broken by the SOI in 2D and 3D systems for the symmetries and materials we investigated.
Even though the damping is finite in the first two cases (as shown in Fig.~\ref{fig:alpha}d), the divergence is still present in the latter (Fig.~\ref{fig:alpha}b).

We can also recognize that the mathematical expression for $\alpha$ in terms of the mean-field susceptibility given in Eq.~\eqref{slopeofchi0withoutSOI} is similar to the conductivity one (i.e., the slope of a response function)~\cite{Bastin:1970jn} --- which leads to the same issues when approaching the clean limit ($\eta \rightarrow 0$).
However, the physical meaning is the exact opposite: While the divergence of the conductivity represents an infinite acceleration of an ideal clean system, infinite damping denotes a magnetic moment that is instantly relaxed in whichever direction it points (as $\ud\VEC{M}/\ud t\rightarrow0$ for $\alpha\rightarrow\infty$) --- i.e., no dynamics~\cite{Kikuchi:1956ed,Mallinson:1987jq}.
This means that a clean 3D spin system is infinitely viscous.
Within the constant broadening model, the divergence of the Gilbert damping can also be seen analytically by comparing Eq.~\eqref{slopeofchi0withoutSOI} with the calculations of the torkance done in Ref.~\onlinecite{Freimuth:2014kq}.
By replacing the torque operator and the current density by the spin lowering and raising operators, respectively, the even contribution (in the magnetization) to the response function vanishes and only the odd one remains.
In this approximation, it is also seen that only the Fermi surface quantities are left, while the Fermi sea does not contribute~\cite{Wimmer:2016hh}.
In the limit of low broadenings, this contribution is shown to diverge as $\eta^{-1}$.
This divergence arises from intraband transitions which are still present in the clean limit, and originate from the finite electronic lifetimes introduced by the constant broadening approximation.

The static limit ($\omega\rightarrow0$) is another reason that many authors considered to be behind the divergent damping behavior~\cite{Kambersky:1976gi,Hankiewicz:2007iy,Garate:2009ic,Freimuth:2017eu}.
This limit is taken in Eq.~\eqref{chiinv} in order to eliminate the contribution of terms nonlinear in frequency from the inverse susceptibility (e.g., inertia effects~\cite{Bhattacharjee:2012if,Thonig:2017jf}).
They can be present in the full microscopic calculation of the susceptibility but are not included in the phenomenological model discussed in Appendix~\ref{appx:llg}.
Adding the quadratic term in frequency leads to an inverse susceptibility given by
\[
\label{inertia}
\IM[\chi^{-+}(\omega)]^{-1} = -\frac{\omega}{2\gamma M}\left(\alpha - \omega I\right)
\]
where $I$ is the off-diagonal element of the moment of inertia tensor~\cite{Bhattacharjee:2012if}.
The fit to the expression linear in frequency then yields an effective $\alpha^\text{eff}(\omega)$.
In the vicinity of the resonance frequency, $\alpha^\text{eff}(\omega_\text{res})=\alpha - \omega_\text{res} I$, which is clearly reduced in comparison to the one obtained in the static limit, $\alpha^\text{eff}(0)=\alpha$.
According to Ref.~\onlinecite{Thonig:2017jf}, $I\sim\alpha/\eta$, which explains the discrepancy between the FMR and the ISM seen in Fig.~\ref{fig:alpha}b as $\eta\rightarrow0$.
We can then conclude that the static limit is not the culprit behind the divergence of $\alpha$ in the clean limit.

\section{Conclusions}
\label{sec:conclusions}

In this work, we presented a study of different methods to calculate the intrinsic Gilbert damping $\alpha$, offering a panorama of how the approaches are related and their range of validity (see Fig.~\ref{fig:diagram}).
They can be grouped into three main categories: the methods that directly employ the results of full microscopic calculations of the dynamical magnetic susceptibility $\chi(\omega)$ (FMR and ISM); 
the exchange-torque method (XC-TCM), which is also based on $\chi(\omega)$ but making use of the mean-field approximation;
and the spin-orbit torque-correlation method (SO-TCM), obtained from the (spin-orbit) torque-torque response via an equation of motion for $\chi(\omega)$.
While the FMR, ISM and XC-TCM include all the contributions to the magnetic relaxation, the SO-TCM provides only the intrinsic contribution due to the angular momentum transfer to the orbital degrees of freedom (not including, for example, the spin pumping mechanism).
The XC- and SO-TCM, given by Eqs.~\eqref{xctcm} and~\eqref{sotcm}, are predominant in the literature due to their simplicity in obtaining $\alpha$ in terms of Fermi level quantities.
It is important to note, however, that they rely on approximations that may not always be fullfilled~\cite{Costa:2015hp}.

In order to implement and compare the different methods, we constructed a unified underlying framework based on a multi-orbital tight-binding Hamiltonian using as case studies the prototypical bulk 3D systems: bcc Fe, fcc Co and fcc Ni.
For this set of materials, the different methods lead to similar results for $\alpha$, showing that the corresponding approximations are well-founded.
Even when the SOI strength is scaled up by one order of magnitude, this excellent agreement remains, as we explicitly verified for bcc Fe.
We found one method that falls out-of-line with the others in the clean limit, namely the perturbative form of the SO-TCM formula~\cite{Edwards:2016be,Umerski:2017cn}.
In this case, although the equation is identical to the well-known Kambersky formula, Eq.~\eqref{sotcm}, the electronic states used to evaluate it do not include SOI.
By comparison with the other methods, we conclude that the results obtained by the perturbative SO-TCM are only valid in the large broadening regime (compared to the SOI strength).
Central to our analysis was a careful study of the convergence of our results with respect to the number of k-points, reaching up to $10^{10}$ k-points in the full Brillouin zone.

The behavior of $\alpha$ is intimately connected with the constant broadening approximation for the electronic lifetimes.
For high temperatures, the Gilbert damping increases with increasing temperature ($\alpha \sim \eta$), while for low temperatures it diverges for 3D ferromagnets ($\alpha \sim 1/\eta$), but not for 2D (ferromagnetic monolayers).
Our calculations revealed that the high temperature values of $\alpha$ arise mostly from the broadening of the electronic states.
In Ref.~\onlinecite{Mankovsky:2013ii}, the strongly increasing behaviour of $\alpha$ for high temperatures was found to be spurious, and cured employing a more realistic treatment of disorder and temperature, and the so-called vertex corrections.
We found that the contribution of the intrinsic SOI to $\alpha$ is additive to the one arising from the broadening, and can be easily extracted by performing a calculation of $\alpha$ without SOI and subtracting this result from the SOI one, $\alpha - \alpha_\text{noSOI}$.
Combined with the ISM, this provides a relatively simple and accurate way to obtain the intrinsic damping, which discounts contributions from the additional broadening $\eta$.
This establishes an alternative way of accessing the high temperature regime of $\alpha$.

The low-temperature divergence of $\alpha$ when approaching the clean limit for 3D ferromagnets has also been the subject of much discussion.
The first difficulty is in establishing numerically whether this quantity actually diverges or not.
Our results consistently show an increase of $\alpha$ with decreasing $\eta$, down to the smallest achievable value of $\eta = \SI{0.14}{\milli\electronvolt}$ (Fig.~\ref{fig:ialpha_kpt}), with no hints of a plateau being reached, but only when accounting for SOI.
This divergence arises from the intraband contributions to $\alpha$, as discussed in Ref.~\onlinecite{Gilmore:2007ev}.
Refs.~\onlinecite{Edwards:2016be,Umerski:2017cn} used perturbation theory arguments to claim that such intraband contributions should be excluded.
However, as we discussed in Sec.~\ref{staticmethods}, adapting the formalism of Ref.~\onlinecite{Freimuth:2014kq} to the calculation of $\alpha$ shows that these intraband terms are enabled by the constant broadening approximation, and so should be included in the calculations.
Contrary to the high temperature regime, works that employ a more realistic treatment of disorder and temperature still find the diverging behavior of $\alpha$~\cite{Mankovsky:2013ii,Mahfouzi:2017kv}.

In real experiments, any kind of material disturbance such as disorder or temperature effects leads to a finite value of the damping.
Besides that, a non-uniform component of the oscillatory magnetic field (either from the apparatus itself or due to its limited penetration into the sample) induces excitations with finite wave vectors and finite linewidths~\cite{Prange:1972kp,Hankiewicz:2008dwa}.
A different way to determine the damping parameter is using the time-resolved Magneto-Optic Kerr Effect (TR-MOKE)~\cite{Bland:2005gx,Neudecker:2006hl}.
It has the advantage that, as it accesses a smaller length scale ($\sim\SI{1}{\micro\meter}$) than FMR experiments (which probe the whole magnetic volume), the measured magnetic properties are more homogeneous and thus the effect of linewidth broadening may be weaker.
The magnetic excitations in nanomagnets can also be probed by recent refinements of FMR experimental setups~\cite{Nembach:2013cn,Casola:2018hq}.

Although the methods we described here are general, we did not explicitly addressed non-local sources of damping such as the spin-pumping~\cite{Tserkovnyak:2002ju}.
As a future project, we plan to ascertain whether our conclusions have to be modified for systems where this mechanism is present.
Systems that combine strong magnetic elements with heavy ones possessing strong SOI are expected to have anisotropic properties, as well-known for the magnetic interactions~\cite{Hellman:2017fr}.
It is then natural to explore when the Gilbert damping can also display significant anisotropy, becoming a tensor instead of a scalar quantity~\cite{Steiauf:2005bv,Thonig:2014kt}.
Indeed, this has been observed experimentally in magnetic thin films~\cite{Baker:2016gx,Chen:2018jw}.
As the SOI, magnetic non-collinearity can also lead to other forms of damping in domain walls and skyrmions~\cite{Yuan:2014en,Jue:2016hg,Akosa:2016ct,Freimuth:2017eu,Kim:2018bq,Mankovsky:2018va}.
From the microscopic point of view, the potential coupling between transverse and longitudinal degrees of freedom allowed by the non-collinear alignment should also be considered.
Lastly, higher order terms in frequency, such as the moment of inertia~\cite{Ciornei:2011ep,Fahnle:2011fv,Bhattacharjee:2012if,Li:2015ht,Thonig:2017jf}, might also become important in the dynamical magnetic susceptibility for large frequencies or for antiferromagnets, for instance.

The description of magnetization dynamics of real materials helps to design new spintronic devices able to control the flow of information.
Our work sheds light on fundamental questions about the main relaxation descriptions used in the literature and sets ground for future theoretical predictions.

\appendix

\section{Ground-state Hamiltonian}
\label{appx:theory}

In this Appendix, we give the explicit forms of the terms in the Hamiltonian written in Eq.~\ref{hamilt}.
As the investigated systems only have one atom in the unit cell, the site indices are omitted.

The electronic hoppings in the lattice are described by
\begin{equation}\label{h0}
\begin{split}
\hat{H}_0 =& \frac{1}{N}\sum_{\VEC{k}\sigma}\sum_{\mu\nu} t_{\mu\nu}(\VEC{k}) c_{\mu\sigma}^{\dag}(\VEC{k})c_{\nu\sigma}(\VEC{k})\quad,
\end{split}
\end{equation}
with $c_{\mu\sigma}^{\dag}(\VEC{k})$ and $c_{\nu\sigma}(\VEC{k})$ being the creation and annihilation operators of electrons with spin $\sigma$ and wave vector $\VEC{k}$ in the orbitals $\mu$ and $\nu$, respectively.
The tight-binding parameters $t_{\mu\nu}(\VEC{k})$ were obtained by fitting paramagnetic band structures from first-principles calculations up to second nearest neighbors~\cite{Papaconstantopoulos:1986hr}, within the two-center approximation~\cite{Slater:1954hi}.

The electron-electron interaction is characterized by a local Hubbard-like~\cite{Hubbard:1963jq} interaction within the Lowde-Windsor approximation~\cite{Lowde:2006ii}, resulting in the mean-field exchange-correlation term
\begin{equation}\label{hxc}
\begin{split}
\hat{H}_\text{xc} 
= -\sum_{\substack{\mu\in d\\ \sigma}}\frac{U}{2}\Bigg\{&M^\alpha\sigma^\alpha_{\sigma\sigma'}+\sum_{\nu\in d}\delta n_{\nu}\left(2\delta_{\sigma\sigma'}\delta_{\mu\nu}-\delta_{\sigma\sigma'}\right)\Bigg\}\\
&c_{\mu\sigma}^{\dag}(\VEC{k})c_{\mu\sigma'}(\VEC{k})\ .
\end{split}
\end{equation}
Here, $U$ is the local effective Coulomb interaction, $M^\alpha$ and $\sigma^\alpha$ are the $\alpha$-component of the magnetic moment vector (summed over the $d$ orbitals) and of the Pauli matrix, respectively.
$\delta n_\mu$ is the change in the occupation of orbital $\mu$ compared to the DFT calculations included in Eq.~\ref{h0}. 
$M^\alpha$ and $\delta n_\mu$ are determined self-consistently. 

The atomic SOI is described by 
\begin{equation}
\begin{split}
\hat{H}^{\text{SOI}} 
=& \lambda \sum_{\substack{\mu\nu\\\sigma\sigma'}}\hat{L}^\alpha_{\mu\nu}\hat{S}^\alpha_{\sigma\sigma'} c_{\mu\sigma}^{\dag}(\VEC{k})c_{\nu\sigma'}(\VEC{k})\quad,
\end{split}
\end{equation}
where $L^\alpha$ and $S^\alpha$ are the $\alpha$ components of the orbital and spin vector operators, respectively.
The strength of the SOI, $\lambda$, is also obtained from first-principles calculations.

The interaction with a static magnetic field $\VEC{B}^\text{ext}$ is described by
\begin{equation}\label{hz}
\begin{split}
\hat{H}_\text{ext} =& B^\alpha_\text{ext}\sum_{\substack{\mu\nu\\\sigma\sigma'}}(\hat{L}^\alpha_{\mu\nu}\delta_{\sigma\sigma'}+\sigma_{\sigma\sigma'}\delta_{\mu\nu})c_{\mu\sigma}^{\dag}(\VEC{k})c_{\mu\sigma'}(\VEC{k})\quad,
\end{split}
\end{equation}
where $\MUB$ is absorbed to $B^\alpha_\text{ext}$ and we used $g_\text{L}=1$ and $g_\text{S}=2$ as the Land\'e factors for the orbital and spin angular momentum.

\section{Phenomenology of FMR}
\label{appx:llg}

The semi-classical description of the magnetization is obtained using the Landau-Lifshitz-Gilbert (LLG) equation \eqref{llg}~\cite{Gilbert:2004gx}.
The effective field acting on the magnetic moment is obtained from the energy functional of the system as $\VEC{B}_\text{eff}(t) =-\partial E/\partial\VEC{M}$.
For the symmetries we investigate, the model energy\cite{VuFind-176410} for the 3D cubic cases\cite{Halilov:1998hs} can be written as
\begin{equation}
\begin{split}
E^\text{3D}(\VEC{M}) = \frac{K_4}{M^4} (M_x^2M_y^2+M_y^2M_z^2+M_x^2M_z^2) - \VEC{M}\cdot\VEC{B}_\text{ext}
\end{split}\ ,
\end{equation}
while for 2D systems,
\begin{equation}
\begin{split}
E^\text{2D}(\VEC{M}) = -\frac{K_2}{M^2} M_z^2 - \VEC{M}\cdot\VEC{B}_\text{ext}
\end{split}\ .
\end{equation}
Positive values of $K_4$ and $K_2$ yield easy magnetization direction along the (001) direction.

We consider magnetic moments pointing along the easy axis, which defines the $\VEC{\hat{z}}$ direction. 
Static magnetic fields are applied along the same orientation.
The magnetic moment is set into small angle precession, $\VEC{M} = M\VEC{\hat{z}} + \delta M_x(t)\VEC{\hat{x}} + \delta M_y(t)\VEC{\hat{y}}$, by an oscillatory field in the transverse plane, i.e., $\textbf{B}_\text{ext}(t) = B_{\text{ext}}\VEC{\hat{z}} +\delta\mathbf{B}_\text{ext}(t)$.
In this form, the effective field (linear in the transverse components of the magnetization) is given by $\VEC{B}_\text{eff}(t) = \VEC{B}_\text{an}(t) + \VEC{B}_\text{ext}(t)$, with
\begin{align}\label{maefields}
\VEC{B}_\text{an}^\text{3D}(t) = -\frac{2K_4}{M^2}(\delta M_x\VEC{\hat{x}} + \delta M_y\VEC{\hat{y}})\text{ , and } \VEC{B}_\text{an}^\text{2D} = \frac{2K_2}{M}\VEC{\hat{z}}
\end{align}
being the anisotropy fields for 3D and 2D systems, respectively.
In the following expressions, $K_4$ and $K_2$ appear in the same way, so they are denoted by $K$.

The Fourier transform of the linearized equation of motion can be written using the circular components $\delta M_\pm = \delta M_x \pm \iu\delta M_y$.
Within this convention, $\delta M_-/\delta B_- = \chi^{-+}/2$ and
\begin{equation}\label{llg+-}
\chi^{-+}(\omega) = \frac{-2\gamma M}{\left[\omega - \gamma(B_{\text{ext}}+B_{\text{an}})\right]-\iu\alpha\omega}\ ,
\end{equation}
where $B_\text{an} = 2K/M$.

\acknowledgments

We are very grateful to R. B. Muniz, A. T. Costa and D. M. Edwards for fruitful discussions. 
The authors also gratefully acknowledge the computing time granted through JARA-HPC on the supercomputers JURECA and JUQUEEN at Forschungszentrum J\"ulich, and the computing resources granted by RWTH Aachen University under project jara0175.
This work is supported by the European Research Council (ERC) under the European Union's Horizon 2020 research and innovation programme (ERC-consolidator grant 681405 -- DYNASORE).

\bibliography{damping.bib}

\end{document}